# Nanoscale domain-wall dynamics in micromagnetic structures with weak perpendicular anisotropy

Tim A. Butcher[1,2,*], Nicholas W. Phillips[1,3], Abraham L. Levitan[1], Markus Weigand[4],
Sebastian Wintz[4], Jörg Raabe[1], and Simone Finizio[1]

[1]*Paul Scherrer Institut, 5232 Villigen PSI, Switzerland*
[2]*Max Born Institute for Nonlinear Optics and Short Pulse Spectroscopy, 12489 Berlin, Germany*
[3]*Minerals Resources CSIRO, 3168 Clayton, Australia*
[4]*Helmholtz-Zentrum Berlin für Materialien und Energie, 14109 Berlin, Germany*

Time-resolved pump-probe soft x-ray ptychography and scanning transmission x-ray microscopy were employed to study the magnetic domain-wall dynamics in microstructures of permalloy ($Ni_{81}Fe_{19}$; Py) with a weak growth-induced perpendicular magnetic anisotropy. The x-ray magnetic circular dichroism images of a micrometer-sized Py square (160 nm thickness) and an elliptical disk (80 nm thickness) show flux-closure patterns with domain walls that fall into alternating out-of-plane (OOP) magnetization states precipitated by the perpendicular anisotropy, which is a precursor of the nucleation of stripe domains at higher thicknesses. An oscillating magnetic field at frequencies from tens of MHz to GHz and up to 4 mT magnitude excited dynamic modes in the domain walls along with the vortex core gyration. The domain-wall dynamics include the translation of inversion points of the OOP magnetization and nucleation of one-dimensional spin waves.

Magnetization dynamics lie at the heart of the technological applications of magnetic materials [1]. Magnetic micro- and nanostructures that feature Landau flux-closure magnetization patterns in order to minimize their stray field [2] present excellent model systems for their study. In the case of micron-sized magnetic squares, the flux-closure pattern consists of four triangular domains with the magnetization in-plane (IP) and arranged head to tail. The domains are separated by domain walls at which the magnetization directions change by 90°. At the center of the square lies a vortex core in which the magnetization points out-of-plane (OOP). Vortices contain two degrees of freedom: the helicity given by the rotation direction of the IP magnetization direction and the OOP orientation of the vortex core (polarity) [3]. The vortex core motion under excitation [4–8] can also create spin waves that are central to the field of magnonics [9,10]. However, the domain walls themselves are also excited by time-dependent magnetic fields and can develop their own dynamics including one-dimensional spin waves [6,11–16].

Although soft magnetic materials have the tendency for the magnetization to lie purely IP, thick films of permalloy ($Ni_{81}Fe_{19}$; Py) reveal a weak perpendicular anisotropy. The columnar granular growth that causes a local OOP shape anisotropy can lead to an OOP reorientation of the magnetization with the formation of stripe domains at a critical thickness [17–24]. Such behavior is general and also observed in other soft magnetic materials [25,26]. The onset of stripe domains can also be triggered by stress in soft magnetic thin films [27–29] and is always strongly dependent on the growth conditions. Originally, there was thought to be a sharp transition from the IP magnetized to stripe domain state at a critical thickness [21]. However, it was shown by magnetic force microscopy that there exists a range of intermediate thicknesses in which a weak perpendicular anisotropy arises that alters the domain-wall structure while the film remains magnetized IP [22,23]. Domain walls represent magnetic inhomogeneities that are energetically susceptible to the formation of a striped pattern of alternating OOP magnetization. Eventually, stripe domains nucleate into the entire film at a critical thickness [23]. Stripe domains are a relevant propagation medium for spin waves [30,31] as are the domain walls [6,11–16]. Thus, the striped domain walls that combine the two entities constitute an intriguing and unexplored system in terms of magnetization dynamics. Imaging magnetic domain-wall dynamics requires a magnetic microscopy technique with both high spatial and subnanosecond temporal resolutions. The following is a study of micron-sized patterned elements of permalloy ($Ni_{81}Fe_{19}$; Py) in the brink state with striped domain walls and their reaction to oscillating magnetic fields. The excitation was imaged with pump-probe soft x-ray microscopy in two configurations: ptychography and scanning transmission x-ray microscopy (STXM).

Microscopy techniques employing synchrotron radiation have been highly successful in tracking magnetization changes by providing sequences of time-resolved images obtained with x-ray magnetic circular dichroism (XMCD) at the x-ray absorption edges of magnetic elements [6–10,16,32–35]. The XMCD contrast is maximized in the soft x-ray energy range, which includes the $L$-edges of the magnetic transition metals. STXM, in which the sample is scanned through a focused x-ray beam, has established itself as one of the most popular methods for imaging magnetization dynamics [9,36,37]. Sufficient temporal bandwidth is typically achieved by using a small area avalanche photodiode (APD), which allows time-resolved imaging of magnetization dynamics at numerous excitation frequencies above and below the repetition rate of the synchrotron source (500 MHz for the Swiss Light Source) [38]. Nonetheless, despite the achievable temporal resolution, which can be pushed beyond the limit imposed by the duration of the x-ray pulses [39], the shortcoming of STXM lies in the spatial resolution. This is dictated by limitations in the fabrication of Fresnel zone plates (FZPs) and is around 25 nm for regular measurements, with a record of 7 nm [40].

---

*Contact author: tim.butcher@psi.ch

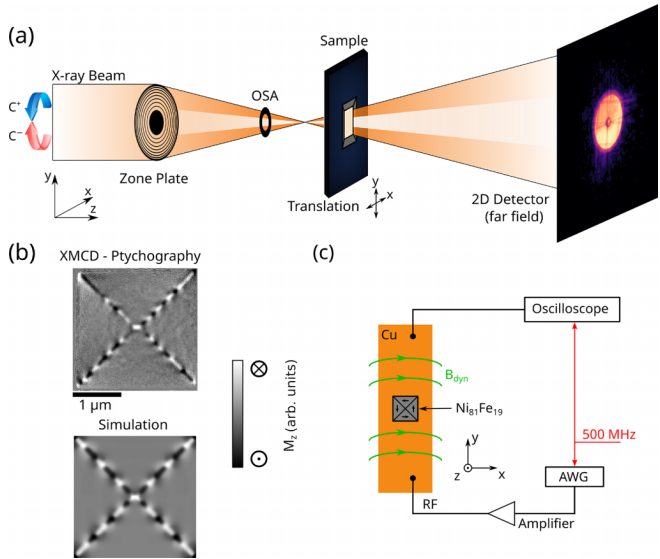

FIG. 1. (a) Soft x-ray ptychography in defocused probe mode. Diffraction patterns from overlapping regions on the sample are captured. (b) XMCD amplitude image of the static Py square from ptychography at the Ni $L_3$-edge and the simulated $M_z$ distribution. Subtraction of the images with opposing x-ray helicities results in the XMCD image. Perpendicular and parallel anisotropies cause striped domain walls and the elongated vortex core, respectively. (c) Sketch of the excitation setup in the pump-probe mode. An rf-current from an arbitrary waveform generator (AWG) causes a periodic Oersted field in the $x$-direction, which is frequency locked to the 500 MHz bunches of the synchrotron.

The coherent diffractive imaging method of soft x-ray ptychography [41] is capable of achieving a spatial resolution that surpasses that of STXM and has been successfully employed for the imaging of ferroic materials including cases of noncollinear magnetism [42–44]. A sketch of the setup can be seen in Fig. 1(a) and the only difference to that of STXM is the detector. Circularly polarized soft x-rays are focused by a FZP and illuminate the sample downstream of the focal plane. An order selecting aperture (OSA) serves to remove both higher orders of x-rays diffracted by the FZP and undiffracted light. The x-rays are scattered by the sample and registered with a two-dimensional pixelated detector. Complex-valued images are reconstructed from the diffraction patterns of overlapping points of the sample that are obtained by scanning it through the beam [45]. Compatible pixelated detectors do not have the necessary bandwidth to resolve individual x-ray bunches and therefore ptychography currently only enables the imaging of dynamics in pump-probe mode with the excitation frequencies locked to an integer multiple of the repetition rate of the synchrotron pulses [46].

The first sample was a thermal evaporation grown Py square with 160 nm thickness and 2.5 μm diameter on an x-ray transparent $Si_3N_4$ membrane. The sample fabrication is described in the Supplemental Material [47]. Ptychographic imaging was performed at the Ni $L_3$-edge (856 eV) and XMCD maps of the OOP magnetization $M_z$ were obtained with circularly polarized x-rays at normal incidence to the sample, which is detailed in the Methods section in the Supplemental Material [47]. The images were reconstructed from the diffraction patterns with the PTYCHOSHELVES software package [48–51] (details in Supplemental Material [47]). The ptychographic XMCD amplitude image of the static Py square is shown in Fig. 1(b), together with the OOP component of the magnetization simulated using the MUMAX$^3$ finite-difference framework [52]. The XMCD images were calculated as the difference of the amplitudes of the ptychographic reconstructions for left- and right-hand circularly polarized illumination. A Landau flux closure pattern is formed in the Py square and the IP domains show no XMCD contrast in this geometry. The entities that have an OOP component are the vortex core and the domain walls that separate the triangular domains.

The thickness of the Py square affects the configuration of the approximately 80 nm-wide domain walls, which were found to have split into different regions of inward and outward magnetization in order to minimize the dipolar energy by flux closure at the sample surface. This is opposed to the common case in which the magnetic moments undergo an OOP rotation in unison akin to a Bloch wall during the 90° reorientation. The explanation lies in the weak perpendicular anisotropy in the 160 nm-thick Py film that is on the verge of flipping the magnetization to an OOP orientation with stripe domains [22,23]. The perpendicular anisotropy originates in the shape anisotropy stemming from the columnar growth of the Py thin film. The weak perpendicular anisotropy diminishes the domain-wall thickness and brings about a sharper change of the OOP magnetization component, which becomes sufficiently sizable for a stray field to appear. This forces the domain walls into regions of alternating magnetization directions for compensation of the stray field before the entire film forms stripe domains at a critical thickness [23]. Hence, the domain walls appear as striped patterns in the XMCD image with OOP sensitivity.

Inspection of the vortex core in the XMCD image shows that it is horizontally elongated with an extent of 120 nm in $x$ and 50 nm in $y$. The reason for this is a slight uniaxial IP anisotropy arising during the growth of the sample [35,53]. The elongation of the vortex core was replicated in micromagnetic simulations by the introduction of an IP component of the anisotropy in $x$-direction ($K_u = 3$ kJ m$^{-3}$). Such asymmetries are often encountered in thin films grown by thermal evaporation, where uniformity is sacrificed for higher deposition rates. Furthermore, the XMCD contrast of the vortex core is not sharp, which evinces a barrel-like magnetic vortex with columnar structure [35]. This three-dimensional modification in which the vortex core diameter is minimized at the surfaces is often observed in thick magnetic microstructures.

The simulated ground state of the Landau flux-closure pattern can adopt the experimentally observed striped domain-wall configuration by setting the perpendicular magnetic anisotropy to $K_u = 8.75$ kJ m$^{-3}$. The simulations also show that the elongation of the vortex core in the *x*-direction by the IP anisotropy modifies the stripes of the domain walls in the top and bottom quadrants of the square. The change in magnetization between different segments of the domain walls is continuous and there is no singularity observed that would signify the presence of Bloch points [54,55]. While the experimentally observed aperiodic domain-wall configuration differs from the ideal theoretical case due to the presence of pinning sites and local variations of the anisotropy that were not accounted for in the simulation, two other experimental observations can be verified: first, the arrow shape of the stripes with a faintly protruding OOP magnetization of the shafts into the domains. Second, the OOP component of the magnetization inverts abruptly in the corners of the square to minimize the stray field.

Subsequently, the dynamics of the flux closure state were investigated. For this, the sample was excited by the radio-frequency (rf) Oersted field from a Cu stripline fabricated on top of the Py square, which is sketched in Fig. 1(c). Frequency locking of the rf-current to harmonics of the probing signal at 500 MHz enabled pump-probe imaging of the dynamics by ptychography with XMCD. Varying the delay of the magnetic excitation with respect to the x-ray pulses allows dynamic imaging with temporal resolutions limited to the x-ray bunch length [approximately 70 ps full width at half maximum (FWHM) at the Swiss Light Source]. The results for rf of 500 and 1000 MHz are displayed in Figs. 2(a) and 2(b), respectively. The corresponding animations can be viewed in Supplemental Material Animations 1 and 2 [47].

Due to the presence of the IP anisotropy, the vortex core moves on an elliptical trajectory with a major and minor axis of approximately 50 and 20 nm, respectively, at 500 MHz. The vortex core motion was accompanied not only by bending of the domain walls but also by excitation modes within them. At 500 MHz, the stretching and contraction of several sections of the domain walls by approximately 200 nm is observed as the domain walls slightly bend during the cycle. This is particularly evident in the bottom right part of the domain walls, in which one segment almost completely merges with its neighbor during the periodic motion around one node. The behavior is also observed at 1000 MHz, but the reaction of the domain walls is more rigid. Above excitations of 1000 MHz and up to the maximum probed frequency of 3000 MHz, the domain walls cannot follow the excitation and the stripe configuration of the domain walls remains unperturbed. No excitation of spin waves was observed in the domains at these frequencies. Although the ground state of the domain wall displayed in Fig. 1(b) is reminiscent of the one-dimensional propagation of spin waves in domain walls [14,16], no frequency or wave vector can be assigned to the movement of the nodes in the domain wall detected here.

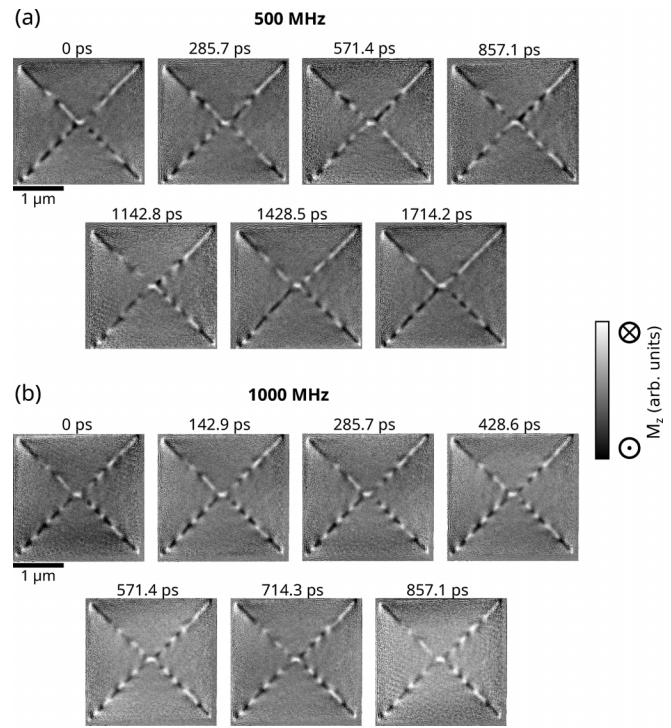

FIG. 2. XMCD ptychographic image sequences of the magnetization configuration upon excitation of the Py square with a 4 mT amplitude in the horizontal direction (Supplemental Material Animations 1 and 2 [47]). Gyration of the vortex core, motion of the domain walls, and their segments are observed. (a) 500 MHz excitation. (b) 1000 MHz excitation.

The detected domain-wall modes were reproduced in micromagnetic simulations of the dynamics at 500 MHz, which are shown in Fig. 3 (Supplemental Material Animations 3 and 4 at 500 and 1000 MHz, respectively [47]). The translation of the domain wall and the pumping of its segments are strongly dependent on the magnitude of the perpendicular magnetic anisotropy. A comparison of the simulations at 500 and 1000 MHz also confirms the increasing rigidity of the system at higher frequencies. The simulations verify that the expansion or shrinking of the stripes in the domain wall of each quadrant of the square can be related to the location of the vortex core during its gyration. The shift of the vortex core into one of the quadrants flips the magnetization direction in the adjacent part of the domain wall to that of the vortex core. This expanding motion is then transmitted along the striped domain wall and reverts when the vortex core exits the respective quadrant.

For this sample, the region swept by the motion of the vortex core does not appear to exhibit "deep" pinning sites, as observed by the fact that the vortex core gyration is smooth. Even so, the presence of pinning sites around the striped domain-wall regions may affect their dynamics, which could explain differences between the experimental observation and the micromamagnetic simulation.

For the purpose of understanding the magnetization changes within the Py square during a 500 MHz excitation, the cross sections of the simulated magnetization distributions for a cut along the domain wall are summarized in

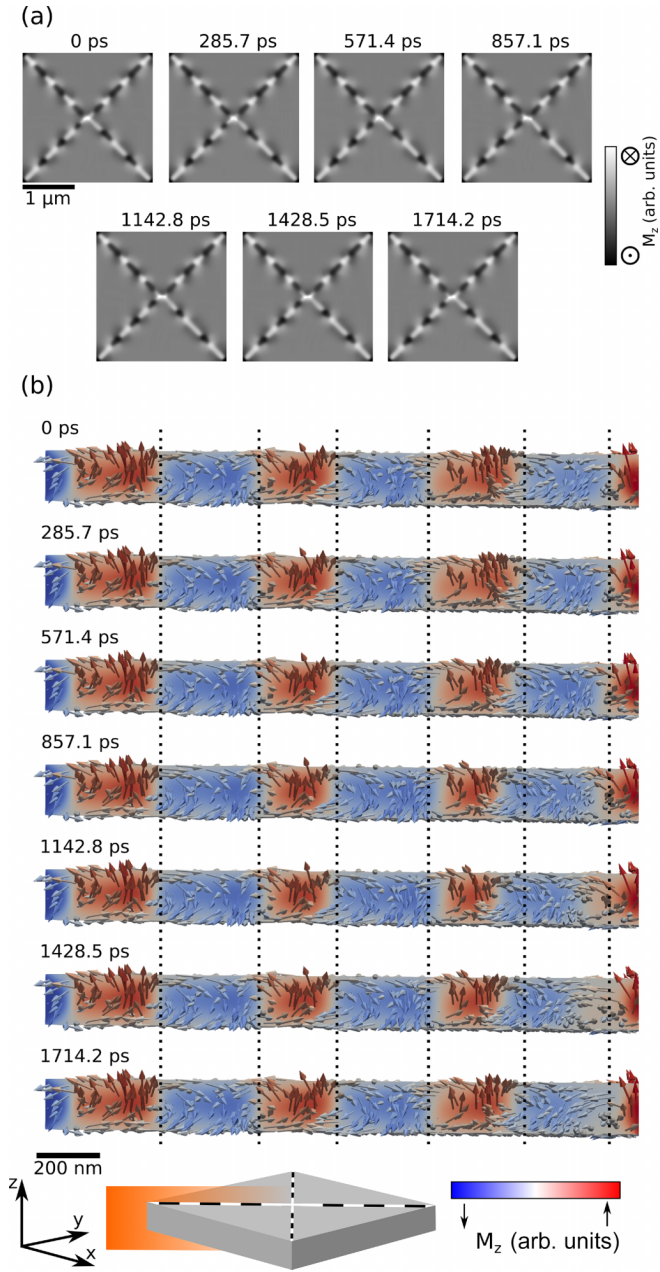

FIG. 3. Micromagnetic simulations of the magnetization dynamics with a 500 MHz excitation (Supplemental Material Animation 3 [47]). (a) The OOP component $M_z$ in the domain walls follows the experimental observation in Fig. 2(a). (b) 2D configuration of **M** in the plane along the domain wall up to the vortex core during 500 MHz excitation. The geometry of the simulated cross section is illustrated below in orange.

Fig. 3(b). The simulations feature an accordionlike motion of the domain-wall segments during the excitation period, in which the up and down regions are alternatingly compressed. Notably, the magnetization curls IP at the surface of the domain-wall segments. These structures are known as Néel caps that are encountered around Bloch domain walls and serve to minimize stray fields emanating from the surface [23,56,57]. In two dimensions, the combination of the Bloch walls and Néel caps resemble vortices that curl around the different sections of the domain wall. The Néel caps move in unison with the 90° Bloch-like domain-wall segments in the studied Landau flux closure pattern (see Supplemental Material Animation 5 [47]). Unlike the experimental results, the different domain-wall segments do not merge completely in the micromagnetic simulations.

In order to investigate the generality of the observed behavior, a second Py sample that was grown by sputter deposition was analyzed. This sample was in the form of an ellipse-shaped disk of 80 nm thickness with axes of 9.5 and 5.3 µm. In order to analyze the response to excitation frequencies different from $N \times 500$ MHz, time-resolved (TR)-STXM with an APD and a field-programmable gate array (FPGA)-based counting board was employed. Note that this frequency range is currently out of reach for ptychography, even though the extension of event-counting Timepix readout devices [58] to soft x-ray detectors [59] may enable the assignment of detection time to incoming photons and dispense with the requirement of frequency locking to the x-ray pulse repetition rate in the future.

Static STXM images of the magnetic domain configuration in the ellipse-shaped Py disk were obtained with circularly polarized x-rays at the Fe $L_3$-edge (nominally 707 eV) and are shown in Figs. 4(a) and 4(b). The IP component was visualized by tilting the sample by 30° with respect to the incoming x-rays [Fig. 4(a)], while the magnetic domain walls were imaged with the x-rays at normal incidence [Fig. 4(b)]. These images show that the Py disk exhibits a flux-closure "eye state," which comprises two adjacent vortices of opposing circulation with their cores located at the corners of a rhombus-shaped central domain [highlighted in Fig. 4(b)]. The domain walls once again decomposed into several segments of alternating OOP magnetization as was the case in the Py square grown by thermal evaporation, although it was only half the thickness [see Fig. 4(b)].

The Py disk was excited by an oscillating magnetic field with an amplitude of approximately 1 mT. The dynamics of the domain structure upon excitation with 71 and 1071 MHz can be seen in Figs. 4(c) and 4(d), respectively. The left images show the domain-wall structure at the beginning of the excitation in a transmission image with a single x-ray helicity, whereas the right two images illustrate the magnetization change with respect to the time-average of an entire cycle at two phases (delay times) of the excitation. The Supplemental Material contains time-sequenced animations of Figs. 4(c) and 4(d) together with the excitations at 2071, 3071, and 4071 MHz (Supplemental Material Animation 6 [47]). The transmission images show that the nodes in the striped domain-wall configuration were affected by the oscillating magnetic field with a homogenization having taken place at 71 MHz.

The two vortex cores in the eye domain structure followed circular trajectories upon excitation. The vortex core gyration radius is more pronounced at low excitation frequencies with a strong response at 71 MHz close to the resonance, which is accompanied by prominent domain-wall motion. The vortex cores on the left and right side react differently with gyration radii of 50 and 100 nm, respectively. What is more, the magnetization of the domain wall is completely flipped half way through the 71 MHz excitation cycle. The corresponding time-sequenced images reveal that the reversal of the magneti-zation emerges from the vortex core during its movement and is transmitted along the domain wall up to certain nodes in the stripe pattern. This shows the impact of the stronger response of the vortex core near resonance on the striped domain-wall state, which is more inclined to flip the magnetization across an extended section of the domain wall until a barrier in form of a node is encountered.

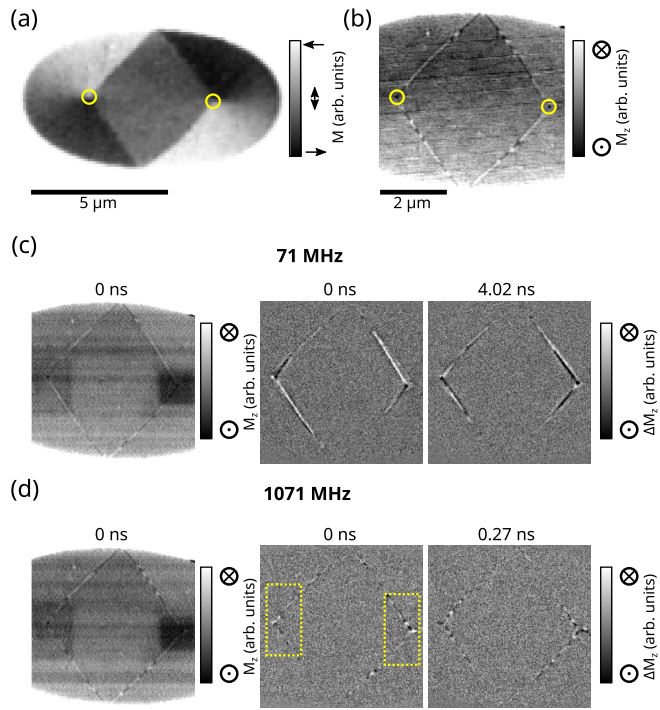

FIG. 4. XMCD STXM images of the ellipse-shaped disk obtained at the Fe $L_3$-edge. (a) IP magnetization component measured with the sample at 30°. (b) $M_z$ measured at normal incidence shows the domain walls and vortex cores, which are highlighted with yellow circles. (c), (d) Pump-probe images of excitation at (c) 71 MHz and (d) 1071 MHz. The corresponding animations and higher excitation frequencies are in Supplemental Material Animation 6 [47]. Left: Transmission images in normal incidence show the domain-wall configuration at the beginning of each excitation cycle. The black squares around the vortex cores are caused by carbon deposition from the focused x-ray beam during further STXM scans. Right: The changes in $M_z$ at phases [0/7] and [2/7] of the excitation cycle are shown. The evolution of a 1D spin wave in the domain wall is uncovered at 1071 MHz within the framed areas.

As was the case in the Py square (Figs. 2 and 3), the nodes in the striped domain walls oscillate at 1071 MHz. Aside from this observation, which is more clear in the transmission images, the magnetization difference images show the emission of one-dimensional spin waves that propagate along the domain walls in the highlighted framed area [16]. At frequencies higher than 1071 MHz, the segments of the domain wall become more static (see Supplemental Material Animation 6 [47]). However, spin waves begin radiating from the individual domain-wall segments into the domains that act as a two-dimensional propagation medium in an arrowlike triangular fashion [60]. When the excitation frequency reaches 6786 MHz, the spiraling vortex core emits an isotropic spin wave (see Supplemental Material Animation 7 [47]). Lastly, a straight 180° domain wall that splits the bidomain Py disk structure can emit plane-wave-like spin waves in a single direction (see Supplemental Material Animation 8 [47]) [16,61].

In conclusion, the magnetic domain-wall dynamics of flux-closure systems with thicknesses on the brink of a change to OOP magnetization were studied by x-ray microscopy. The onset of growth-induced perpendicular magnetic anisotropy divides the 90° Bloch-like domain walls into several segments of alternating magnetization. A complete switch to OOP magnetization at greater thicknesses is accompanied by the emergence of stripe domains throughout the Py [17–24] and the observed vortexlike pattern in the domain wall presages these [22,23]. The individual components of the domain wall stretch and shrink under rf-excitation when the vortex core gyrates, which was observed experimentally in the pump-probe scheme as well as in micromagnetic simulations. The structure of the brink state domain walls and their dynamics depends strongly on the chosen growth mechanism and geometry of the nanostructure. The thinner sputter-deposited Py disk was shown to be less rigid than the thicker square obtained by thermal evaporation. The former allowed observation of the distinct phenomenon of the excitation of one-dimensional spin waves in the domain wall. These coexisted with the movement of nodes in the striped domain walls. The uncovered dynamics in the presence of a weak perpendicular anisotropy point to complex interactions that are of high relevance for magnetic domain-wall motion.

Soft x-ray ptychography measurements were performed at the Surface/Interface Microscopy (SIM-X11MA) beamline of the Swiss Light Source (SLS), Paul Scherrer Institut, Villigen PSI, Switzerland. STXM measurements were carried out at the MAXYMUS end station [62] at the BESSY II electron storage ring operated by the Helmholtz-Zentrum Berlin für Materialien und Energie. We thank the Helmholtz-Zentrum Berlin for the allocation of synchrotron radiation beamtime. Support from the Nanofabrication Facilities Rossendorf (NanoFaRo) at the Ion Beam Center (IBC) of Helmholtz-Zentrum Dresden-Rossendorf (HZDR) is acknowledged. T.A.B. acknowledges funding from the Swiss Nanoscience Institute (SNI) and the European Regional Development Fund (ERDF). N.W.P. and A.L.L. received funding from the European Union's Horizon 2020 research and innovation programme under the Marie Skłodowska-Curie Grant Agreement No. 884104 (PSI-FELLOW-III-3i). We thank E. Fröjdh, F. Baruffaldi, M. Carulla, J. Zhang, and A. Bergamaschi for the development of the LGAD Eiger detector. The LGAD sensors were fabricated at Fondazione Bruno Kessler (Trento, Italy). We thank M. Langer and S. Mayr for preliminary measurements that proceeded the ptychographic imaging in this work. We thank Roland Mattheis (IPHT Jena) for the NiFe sputter deposition of the sample for the STXM measurements.

The data that support the findings of this article are openly available [63].